\documentclass[aps,prd,a4paper,uperscriptaddress,showpacs,
twocolumn,showkeys,amsfonts,amsmath]{revtex4}
\usepackage{times}
\usepackage{epsfig}
\usepackage{psfrag}
\usepackage{graphicx}
\numberwithin{equation}{section}

\begin{document}
\title{Thermodynamics and gravitational collapse}

\author{Daniele Malafarina} \email{daniele.malafarina@polimi.it}
\affiliation{Tata Institute of Fundamental Research
Homi Bhabha road, Mumbai 400005, India}
\author{Pankaj S. Joshi} \email{psj@tifr.res.in}
\affiliation{Tata Institute of Fundamental Research Homi
Bhabha road, Mumbai 400005, India}


\begin{abstract}
It is known now that a typical gravitational collapse
in general relativity, evolving from regular initial data
and under physically reasonable conditions would end in
either a black hole or a naked singularity final state.
An important question that needs to be answered in this
connection is, whether the analogues of the laws of thermodynamics,
as formulated for relativistic horizons are respected by
the dynamical spacetimes for collapse that end in the
formation of a naked singularity.
We investigate here the thermodynamical behaviour of the
dynamical horizons that form in spherically symmetric gravitational
collapse and we show that the first and second laws of black
hole thermodynamics, as extended to dynamical spacetimes
in a suitable manner, are not violated whether the collapse
ends in a black hole or a naked singularity. We then make
a distinction between the naked singularities that result
from gravitational collapse, and those that exist in
solutions of Einstein equations in vacuum axially symmetric
and stationary spacetimes, and discuss their connection
with thermodynamics in view of the cosmic censorship
conjecture and the validity of the third law of
black hole mechanics.
\end{abstract}
\maketitle

\section{Introduction}

The thermodynamical behaviour of black holes in Einstein
gravity was first noted by Hawking, Carter, Bardeen and Bekenstein
in a series of papers, that established what are known now
as the laws of black hole thermodynamics
\cite{therm}.
These laws link the area and surface gravity of the
event horizon of a stationary black hole in classical
general relativity to the entropy and temperature of a
thermodynamical system at equilibrium.
The connection between such diverse concepts
as that of the gravity and the thermodynamical laws
gained physical respectability after the discovery of
Hawking radiation, which, by considering quantum corrections
in a semi-classical context, showed how the analogy was
deeper than that at a formal level, thus proving that black
holes must indeed be considered as black bodies, radiating
at the Hawking temperature
\cite{Hawking}.

Since the first formulation of the laws of black hole
thermodynamics, a lot of effort has been devoted in exploring
the connection between gravity and thermodynamics in the
hope that the analogy can be extended to the whole theory,
making it valid not only on the horizons, and therefore
showing possibly how gravity could be considered as
a thermodynamical theory
\cite{Pad}.

A key issue in this respect is the role played by
the gravitational collapse. The dynamical physical process
that would give rise to a black hole in nature would be
typically the gravitational collapse of a massive star
which shrinks under the force of its own gravity at
the end of its life cycle. In general, a matter of considerable
interest would be the validity of thermodynamical laws
during the dynamical gravitational evolutions of physical
systems in the universe. As it is known now, such a collapse
would terminate in either a black hole or a naked singularity,
depending on the nature of the initial data from which
the collapse evolves, under physically reasonable conditions.
It is known that naked singularities do typically arise in
solutions of Einstein field equations in a wide variety
of situations and one would like to know whether these
spacetimes exhibit a behaviour that is in accordance with
the laws of black hole mechanics.

In stationary or static spacetimes,  we already
know that
extremal solutions (such as in the Kerr geometry or
in the Reissner-Nordstrom spacetime) can lead to the
appearance of naked singularities, and a lot of investigation
has been carried out in order to show whether or not
these manifolds, once interpreted in the context
of thermodynamical gravity, exhibit a behaviour
in accordance with the laws of thermodynamics
\cite{Davies}.
It is important, however, to notice that since a spacetime
singularity, where the densities, curvature, and all physical
quantities diverge,
is not part of the spacetime, understanding thermodynamical
properties of the manifold in the limit of approach to the singularity
could imply at times some technical difficulties.
The eventuality of the occurrence of naked singularities
in the physical universe is a well-known theoretical open
problem that is usually referred to as the cosmic censorship
conjecture (CCC)
\cite{Penrose}.
It is not difficult to see how the validity of
CCC and thermodynamics of naked singular spacetimes are
closely related issues. Therefore the thermodynamical behaviour
of spacetimes with naked singularities has been connected
some times to the possibility to prove the CCC.

It is clear that in order to provide a wider
perspective on the thermodynamical properties of gravity
one has to include dynamical situations, thus extending
the laws of black hole thermodynamics to non-stationary
spacetimes. To this aim, the equivalence of Einstein equations
and the first law of thermodynamics was shown by Jacobson
for all local causal Rindler observers, provided that the
Clausius relation holds
\cite{Jacobson}.
Further, Hayward showed how the laws of black holes
thermodynamics can be extended to trapping horizons in
dynamical spacetimes
\cite{Hayward}.
This in turn has led to the investigation of
dynamical solutions such as the Friedmann-Robertson-Walker
cosmological models in order to establish the connection
between gravity and thermodynamics on a cosmological level
\cite{CaiCao}.

Even then, very little is known as of today regarding
the connection between the thermodynamic behavior and
gravity, for general dynamical asymptotically flat spacetimes
which describe isolated bodies not in equilibrium \cite{ash},
such as for example, the processes that lead to the formation
of a black hole as the endstate of gravitational collapse.
We do know that naked singularities as well as black holes
can form as the endstate of collapse in a wide variety of
models. These naked singularities of collapse are entirely
different from the extremal stationary ones mentioned
above, and still their thermodynamical behaviour is
not fully understood.

Our purpose here is to consider dynamical spacetimes
involving generic spherical gravitational collapse, where
the final state could be either a black hole or a naked singularity.
As mentioned, this is an entirely different issue as
compared to naked singularities that can be found in stationary
spacetimes such as the Kerr-Newman solutions. We shall
therefore investigate the thermodynamical behaviour of the
apparent horizon that develops inside the collapsing cloud
in the limit of approach to the singularity.

Such a study has some resemblance with the analysis
of the thermodynamical properties of the cosmological horizon
in the FRW models, since the homogeneous Lemaitre-Tolman-Bondi
(LTB) model with a non-zero pressure is analogous to the FRW
cosmological model with the time reversed. Still there are
some key differences, for example in the structure of the
trapped surfaces, that in collapse must consider the contribution
from the outer portion of the spacetime. Further the
general collapse situation allows for inhomogeneities and
anisotropic pressures in the collapsing cloud to be considered.
If we allow for inhomogeneities, the LTB collapse scenario
exhibits very different behaviours depending on the properties
of the infalling matter.

We shall show here that, in general the solutions where
the spherical gravitational collapse could lead to naked
singularities, do not violate the analogue of the laws
of thermodynamics for dynamical spacetimes.
In this light, it is important therefore to stress that
such naked singularities of collapse do not exhibit the same
causal structure behaviour as seen in those arising in
stationary cases such as the extremal Kerr naked
singularity.
Furthermore, it has been noted many times
how a theory of quantum gravity might smoothen the behaviour
of quantities approaching the singularity, thus removing
the issue of diverging matter densities. Therefore, a possible
connection of thermodynamics with gravity on a completely
general level would not need to imply that such singularities
cannot occur, while it might rule out other kinds of naked
singularities in case a relation between some formulations
of the CCC and thermodynamics can be proven.

The plan for the paper is as follows. In Section \ref{collapse}
we shall briefly outline the most important features of generic
spherical gravitational collapse and divide the possible final
outcomes into two classes.
These are black holes, in which the singularity that forms at
the end of collapse is hidden within an horizon at all times,
and the naked singularities, in which case the light rays can
escape the singularity at the instant of its formation to
reach faraway observers. In section \ref{thermodynamics},
we study the thermodynamical behaviour of such solutions of
Einstein field equations in order to show how the analogy between
gravity and the first two laws of thermodynamics holds 
regardless of the fact that the final outcome of collapse is a black 
hole or a naked singularity, while in section \ref{third} we briefly 
discuss the third law.
Finally, in Section \ref{discussion}, we will discuss the above
results in view of a possible distinction between different
types of naked singularities, mainly the ones that arise
in static or stationary solutions of Einstein equations, and
those that occur in dynamical collapse models.

\section{Gravitational Collapse} \label{collapse}

Consider a spherically symmetric spacetime, depending on
three metric functions of the comoving time $t$ and the radius
$r$, as given by,
\begin{equation}\label{metric}
    ds^2=h_{ab}dx^adx^b+R(r,t)^2d\Omega^2 \; ,
\end{equation}
with $h_{ab}=\textrm{diag}(-e^{2\nu(t, r)},e^{2\psi(t, r)})$,
$(a,b)=0,1$ and $x_0=t$, $x_1=r$.
The metric functions $\nu$, $\psi$, and $R$ are related
to the energy momentum tensor which in general is
defined by,
\begin{equation}
T_t^t=-\rho; \; T_r^r=p_r; \; T_\theta^\theta=T_\phi^\phi=p_\theta,
\end{equation}
via the Einstein equations.
It is the set of equations that, once a choice
for the initial data is made, determines the final
outcome of gravitational collapse in terms of either a
black hole or naked singularity.

We can define the Misner-Sharp mass of the system
as,
\begin{equation}\label{Misner}
    2E=R(1-G+H) \; ,
\end{equation}
where
\begin{eqnarray}
G(t, r)&=&e^{-2\psi}R'^2 \; , \\
H(t, r)&=&e^{-2\nu}\dot{R}^2\; .
\end{eqnarray}
Then the Einstein equations for the $(0,0)$
component and for the $(1,1)$ component are given as,
\begin{eqnarray}\label{rho}
\rho&=&2\frac{E'}{R^2R'} \; , \\ \label{p}
p_r&=&-2\frac{\dot{E}}{R^2\dot{R}} \; ,
\end{eqnarray}
and these determine the radial pressure and
the energy density as functions of the physical radius
$R$ and of the Misner-Sharp mass.
Together with the other two Einstein equations, namely
\begin{eqnarray}\label{nu}
\nu'&=&2\frac{p_\theta-p_r}{\rho+p_r}\frac{R'}{R}-
\frac{p_r'}{\rho+p_r} \; ,\\ \label{G}
2\dot{R}'&=&R'\frac{\dot{G}}{G}+\dot{R}\frac{H'}{H} \; ,
\end{eqnarray}
they fully determine the structure of the spacetime
\cite{Joshibook}.

Since the Einstein equations do not carry any
information regarding the physical properties of the
matter sources, in order for the solution to be physically
reasonable, some energy and regularity conditions must be
satisfied by the energy momentum tensor, and therefore as
a consequence, by the metric functions. In particular,
the weak energy condition requires positivity of
$\rho$, $\rho+p_r$ and $\rho+p_\theta$, regularity at
the center of the cloud requires that $E(r,t)$ goes
like $r^3$ near the center, and also it is possible to
prove that the pressures must behave like a perfect
fluid near the center, namely we must have $p_r=p_\theta$
in the limit of $r$ approaching zero.

\subsection{Collapse final states}

\begin{figure}[hh]
\includegraphics[scale=1]{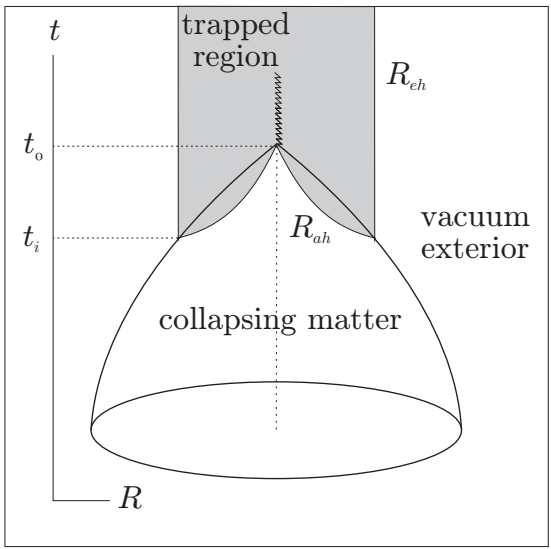}
\caption{Collapse leading to a black hole: The trapped
surfaces form at a time $t_i$ before the formation of the
singularity. The apparent horizon moves inwards from
$R_{eh}$ at $t_i$ to reach zero at $t_0$.}\label{Fig1}
\end{figure}

The complete gravitational collapse of such a
matter distribution generally ends in the formation of
a spacetime singularity, which is indicated by the
divergence of the curvature and the energy density $\rho$.
The possible outcomes of the complete collapse, in
terms of either a black hole or naked singularity are then
characterized by the occurrence and behaviour of
the trapped surfaces developing in the spacetime
as the collapse progresses. In the black hole scenario,
the apparent horizon forms at an outer shell of the
collapsing matter at a stage earlier than the singularity.
The outside event horizon then entirely covers the
final stages of collapse when the singularity forms,
while the apparent horizon inside the matter evolves
from the outer shell to reach the singularity at
the instant of its formation
(see figure \ref{Fig1}).

\begin{figure}[hh]
\includegraphics[scale=1]{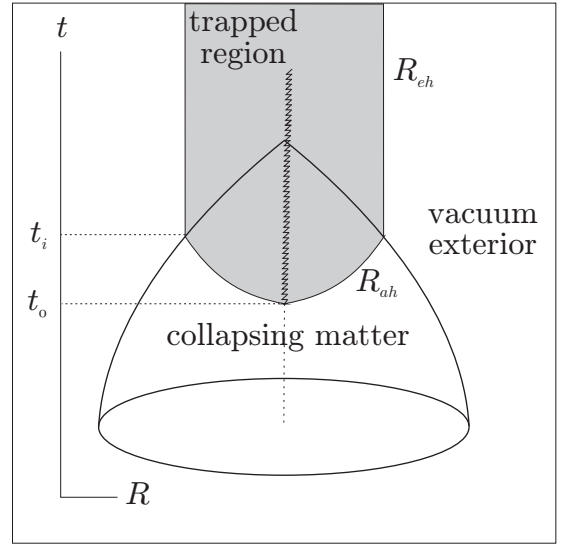}
\caption{Collapse leading to a naked singularity: The
trapped surfaces form at the same time of formation of the
singularity, $t_0$. The apparent horizon moves outwards
from a zero physical radius at $t_0$ to reach a value
$R_{eh}$ at $t_i$.}\label{Fig2}
\end{figure}

In the naked singularity scenario on the other hand,
the trapped surfaces form at the center of the cloud
at the time of formation of the singularity. The
apparent horizon then moves outwards to meet the event
horizon at the boundary of the cloud at a time later
than the time at which the singularity formed. The
instant of formation of the singularity is therefore
not covered by the horizon and it can be shown that
families of light rays and particles can escape the
central singularity and propagate to faraway
observers in the spacetime
(see figure \ref{Fig2}).

Mathematically, it can be proven that in the latter
case a set of future directed null geodesics can emanate
from the singularity to reach faraway observers in the
spacetime
\cite{ndim}.
Physically this means that the ultra-dense region
surrounding the singularity, where the quantum-gravitational
effects would become dominant, can communicate with
the outside universe. We shall note that a wider variety
of scenarios is possible even when non-zero pressures are
included in the collapse analysis. These include
situations where the collapsing matter is radiated away
during the collapse process, and the singularity curve
remains uncovered for a longer time as the trapped surfaces
form later than the singularity
\cite{Mandar}.
Nevertheless for our purposes here, the schematic
view outlined above is enough to distinguish the black hole
case from the naked singularity in gravitational collapse.

Whether naked singularities can actually arise from
physically viable processes has been a matter of much
debate currently and for past many years. The CCC, as
formulated by Penrose, claims that all such occurrences
must be covered at all times by an horizon, ruling out
the possibility for singularities to be visible to
faraway observers. Nevertheless, many counterexamples in
models of collapse exhibit the behaviour described above,
thus suggesting that naked singularities can indeed
form as endstates of complete gravitational collapse of
massive bodies
\cite{Joshi}.
Supporters of the CCC have therefore tried to
invoke the laws of thermodynamics in order to construct
a mechanism by which naked singularities cannot form.
While such claims can in principle be legitimate in some
cases, we shall see that the behaviour of horizons in
gravitational collapse in general does not violate the laws of
thermodynamics, regardless of the fact that the final
outcome is a black hole or a naked singularity.

\section{Thermodynamics} \label{thermodynamics}

The laws of black hole thermodynamics as stated by
Hawking, Carter and Bardeen establish the connection
between gravity and thermodynamics on the event horizon
of a stationary black hole. In the case of vanishing
charge and angular momentum (namely for the Schwarzschild black
hole) these can be written as:
\begin{itemize}
  \item[-] 0th law: $\kappa=const.$ on the horizon.
  \item[-] 1st law: $dM=\frac{\kappa}{8\pi}dA$ on the horizon.
  \item[-] 2nd law: $\frac{dA}{dt}\geq 0$ on the horizon.
\end{itemize}
In the above, $\kappa$ is the surface gravity on
the event horizon, $M$ is the total mass of the system
(corresponding to the Schwarzschild parameter) and $A$ is
the area of the event horizon. The full analogy with
equilibrium thermodynamics is then established once we relate
the area to the entropy $S$ (in geometrized units) via
\begin{equation}
    S=\frac{A}{4} \; ,
\end{equation}
and the surface gravity to the temperature $T$ via
\begin{equation}\label{temp}
    T=\frac{\kappa}{2\pi} \; .
\end{equation}
The generalization to rotating and charged black holes
is possible, but not necessary in view of the present analysis
of collapse, which does not consider the charge and angular
momentum parameters.

\subsection{Trapping horizons}

Typically, the laws of black hole thermodynamics are
defined on the event horizons in vacuum stationary space-times,
like the Schwarzschild or Kerr-Newman solutions, since,
in general, if the metric depends on $t$ we cannot then
find a preferred time direction in the form of a Killing vector.
Therefore we cannot define a temperature in the sense of
Hawking and Bekenstein, since it involves evaluating the
surface gravity on a Killing horizon.

It is nevertheless possible to generalize the results
of black hole thermodynamics to dynamical spacetimes where
the metric components depend on $t$, if one makes use of
the Kodama vector, instead of the Killing vector, to identify
a preferred time direction. Then the thermodynamical
quantities are defined on trapping horizons instead of
the Killing horizons
\cite{Kodama}.

In this context, we are not anymore considering
the laws of black hole thermodynamics at equilibrium. We are
instead considering a dynamical system approaching
equilibrium. It is important therefore to specify clearly
what constitutes the system under consideration. In the
case of gravitational collapse, the natural choice to make
is that of the trapped region that develops as the
collapse evolves, namely the volume of spacetime region
causally disconnected from the outside universe, the
boundary of this region being given by the trapping
horizon. Of course, in the limit of approach to
equilibrium the boundary of the trapped region
approaches the usual event horizon.

At first, as matter starts collapsing under the
effect of gravity, no portions of the spacetime are
trapped. As certain high densities are reached, the trapped
surfaces form and a trapped region develops in the
spacetime. It is this part of the spacetime that evolves
eventually forming the final black hole, possibly in a
static or stationary configuration. Before it settles
to its final state, the boundary of the trapped region is
marked by the presence of an apparent horizon. As we have
seen in the previous section, the apparent horizon typically
develops between the time of formation of the singularity
and the time at which it meets the outer Schwarzschild
event horizon, and the singularity can be either causally
connected or disconnected from the outside universe,
which is decided by the pattern of trapped surfaces
formation as the collapse evolves.

A sphere of physical radius $R$ is said to be trapped,
marginally trapped or untrapped if $h^{ab}\partial_a R \partial_b R$
is smaller than zero, equal to zero, or greater than
zero, respectively. Defining the double null coordinates
$\{\xi^+,\xi^-\}$ in such a way that
\begin{equation}
    ds^2=2d\xi^+d\xi^-+R^2d\Omega^2 \; ,
\end{equation}
where the double null vectors are given by
\begin{equation}
    \partial_\pm=\frac{\partial}{\partial\xi^\pm}
=-\sqrt{2}\left(\frac{\sqrt{G}}{R'}\partial_r\mp e^{-\nu}\partial_t\right) \; ,
\end{equation}
 we can write the expansion of the null geodesic
congruences as,
\begin{equation}
    \theta_\pm=2\frac{\partial_\pm R}{R} \; ,
\end{equation}
from which we can see that the condition
$\theta_\pm>0$ ($<0$) on a sphere of radius $R$ is
enough to ensure that the light rays on the sphere
diverge (converge). It can be shown that if $\theta_+\theta_-$
has non-vanishing derivatives then the spacetime can
be divided into a trapped region and an untrapped region,
separated by a marginally trapped surface.

Defining the trapping horizon as the closure of
the hypersurface obtained through the foliation of
marginal spheres in the whole spacetime, we get the
condition that $R$ must satisfy at all times in order
to describe a trapping horizon, which is given by,
\begin{equation}
    h^{ab}\partial_aR\partial_bR=0 \; .
\end{equation}
In the above double null foliation, supposing
the horizon is given by $\partial_+R=0$, then it is
said to be future if $\partial_-R<0$ (past, if positive),
and outer if $\partial_-\partial_+R<0$ (inner, if positive).
A black hole is defined as a future, outer trapping horizon.
In the case of the apparent horizon, the requirement
that the horizon be `outer' can be dropped. In fact in
the FRW models the horizon is considered to be future but
negativity of the surface gravity indicates that it is an
inner horizon, nevertheless, it is possible to prove
the thermodynamical behaviour of such an horizon.
The same reasoning holds for gravitational collapse.

We see immediately that in the general situation
described by collapse, the trapped horizon in the collapsing
matter is given by the apparent horizon $R_{ah}$ and it
is possible to construct the foliation such that the
apparent horizon is a future horizon.

Nevertheless, the apparent horizons in the situations
for black hole and naked singularity formation are different
(as well as they differ from the FRW cosmological scenario).
In the black hole case, we have $\dot{R}_{ah}<0$ and the horizon
equation is given by $\partial_-R=0$, while in the naked singularity
case we have $\dot{R}_{ah}>0$, and the horizon equation
is given by $\partial_+R=0$. By using the Misner-Sharp mass
it is easy to check that in both cases we are dealing
with a future horizon which is implicitly defined by $2E=R$.

\subsection{Zeroth law}

In the case of dynamical spacetimes the zeroth law
can be formulated by the statement that the total trapping
gravity of a future outer marginal sphere has an upper
bound
\cite{Hayward}.
From this definition it is possible to retrieve
the case of event horizons in stationary spacetimes
considered by Hawking.
It has also been noted that the surface gravity in
dynamical spacetimes can be defined in many different ways
\cite{Nielsen}.
Following Heyward, we can evaluate the surface gravity
on the trapping horizons as,
\begin{equation}\label{surface-gravity}
    \kappa=\frac{1}{2\sqrt{-h}}\partial_a(\sqrt{-h}h^{ab}\partial_bR) \; ,
\end{equation}
which in our case becomes
\begin{equation}\label{k}
    \kappa=\frac{E}{R^2}+\frac{1}{4}(p_r-\rho)R \; ,
\end{equation}
and we can relate the surface gravity of the
apparent horizon with the temperature through equation
\eqref{temp}. It is easy to check that for the event
horizon, the equation \eqref{k} reduces to the
well-known formula for Schwarzschild.

\subsection{Unified first law}

The Einstein equations \eqref{rho} and \eqref{p}
imply that we can write the variation of the Misner-Sharp
mass as
\begin{equation}\label{1st}
    dE=A\Psi+WdV \; ,
\end{equation}
where the area and volume of the sphere are the
geometrical invariants used to define $R$ from
\begin{equation}
    A=4\pi R^2, \; V=\frac{4}{3}\pi R^3 \; ,
\end{equation}
$W$ is the work defined by
\begin{equation}
    W=-\frac{1}{2}h_{ab}T^{ab} \; ,
\end{equation}
and $\Psi$ is the energy-supply defined by
\begin{equation}
    \Psi_a=T^b_a\partial_b R+W\partial_a R \; .
\end{equation}
Equation \eqref{1st} is the Unified First Law, and
it naturally descends from Einstein equations in
spherical symmetry. In fact by differentiating equation
\eqref{Misner} it is easy to show that equation
\eqref{1st} is equivalent to equations
\eqref{p} and \eqref{rho}.
Evaluating equation \eqref{1st} on a trapping
horizon gives the first law of black hole thermodynamics
which states,
\begin{equation}
    \langle dE, z\rangle = \frac{\kappa}{8\pi}
\langle dA, z\rangle + W \langle dV, z\rangle \; ,
\end{equation}
for some vector $z$ tangent to the trapping horizon.
As it has been noted, $z=z^+\partial_++z^-\partial_-$
is not an arbitrary vector but it must satisfy a certain
condition on the horizon. Namely, if the horizon is given
by $\partial_-R=0$ then,
\begin{equation}
    \frac{z^-}{z^+}=-\frac{\partial_+\partial_+R}{\partial_-\partial_+R} \; .
\end{equation}

\subsection{The Clausius relation}

In order to prove that the Unified First Law is
equivalent to the first law of black hole thermodynamics
on the apparent horizon, one has to prove the Clausius
relation,
\begin{equation}
    \langle A\Psi, z\rangle= \frac{\kappa}{8\pi}\langle dA, z\rangle \; ,
\end{equation}
where the left hand side is just the heat flow
$\delta Q$ as defined by the energy momentum tensor,
while the right hand side has the form $TdS$ once
the above mentioned identifications between $T$ and $\kappa$
and $S$ and $A$ are made. We can see then that if the
first law holds on the horizon, the Unified First Law
implies the Clausius relation $\delta Q=TdS$.

\subsection{The Second law}

We shall consider here the second law of black hole
mechanics for dynamical horizons in gravitational collapse,
which states that under the satisfaction of energy conditions
the area of a future (outer) horizon is non-decreasing.

Nevertheless, there are examples in cosmological models,
violating energy conditions but still used to describe the
dark energy in the universe, where the second law at the
horizon is violated. Generally one resorts to the formulation
of a Generalized Second Law of black hole thermodynamics
which states that the sum of the entropy of the horizon
and the entropy of the matter bounded by the horizon (in the
case of a cosmological horizon) does not decrease in time.
These considerations suggest that one must be careful when
talking about entropy in dynamical spacetimes, since a proper
thermodynamical definition of the same, based on the
microscopic behaviour of the spacetime, is still missing.

In the case of gravitational collapse, we would not
consider the entropy of the infalling matter and therefore
we will restrict ourselves to a genuine second law evaluated
on horizons.
For this reason we will neglect considerations regarding the
microstates, as it will be sufficient for us to extend the usual
definition of entropy as the area of the horizon, from the
static case of the event horizon to the dynamical case of
the apparent horizon.
As we have seen, the apparent horizon in the
black hole formation scenario in collapse is not an outer
horizon, and therefore the area theorem does not hold in
this case. Nevertheless we can formulate the second law of
black hole thermodynamics once we consider the trapping
horizon of the spacetime as the union of the inner apparent
horizon and the outer event horizon, thus considering the
horizons as the actual boundary of the trapped region in
the spacetime. It is straightforward then to see that the
volume of the trapped region is increasing in time, both
in the case of the black hole and naked singularity.
As said, in analogy with the static case, we shall define
the entropy of the trapping horizon at any given time as
the derivative with respect to $R$ of the volume of the
trapped region. Therefore, in the black hole case we obtain,
\begin{equation}
    S_h=\pi(R_{eh}^2-R_{ah}^2) \; ,
\end{equation}
where $R_{eh}$ is the Schwarzschild radius in
the exterior spacetime and $R_{ah}$ goes to zero as
$t$ goes to $t_0$. Note that with this definition the
apparent horizon has a negative entropy, in accordance
with the entropy of the horizon in FRW models, but
the entropy of the whole system is positive. In this
case, the horizon forms at $t_i<t_0$ and becomes $R=2M$
for $t\geq t_0$, with $M=E(r_b, t)$ being the total
mass of the system.

In the naked singularity case we obtain,
\begin{equation}
    S_h= \begin{cases}
  \pi R_{ah}^2 & \text{for $t\in[t_0,t_i)$} \\
  \pi R_{eh}^2 & \text{for $t\in[t_i, +\infty)$} \; ,
\end{cases}
\end{equation}
where again $R_{eh}$ is the Schwarzschild radius in
the exterior spacetime and $R_{ah}$ goes from zero at the
time of the formation of the horizon $t_0$ to $R_{eh}$ at
the time $t_i$, when all the matter is bounded by the
horizon. In this case, the horizon forms at $t_0<t_i$
and reaches the value $R=2M$ for $t\geq t_i$.

In both these cases, the apparent horizon curve
$r_{ah}(t)$ is determined implicitly by,
\begin{equation}
    2E(r_{ah}(t),t)=R(r_{ah}(t),t) \; ,
\end{equation}
and in both cases of either the black hole or naked
singularity formation, the horizon settles to the Schwarzschild
event horizon once the collapse ends, at which point we regain
the usual static definition of entropy.
It is straightforward to check that
\begin{equation}
    \frac{dS_h}{dt}\geq 0
\end{equation}
in both cases, and thus, if the above definition of the
entropy is valid from a microscopical point of view, the
second law of thermodynamics holds for the trapping
horizons in the gravitational collapse scenario.

\section{The third law and cosmic censorship } \label{third}

The analogy between black hole mechanics and
thermodynamics would be complete if a satisfactory formulation
of the third law could be given. Unfortunately the status
of the third law regarding black hole mechanics remains
unclear at the present juncture.

There exist two alternative formulations of the
third law of thermodynamics. The `strong' form says that
the entropy $S$ of a system goes to a constant value,
which can be taken to be zero, as the temperature of the
system goes to zero. A weaker statement says that it is impossible
to reach $T$ equal zero through any finite series of physical processes
\cite{Huang}.

When applied to black hole mechanics, it is easy to
find counterexamples to the analogue of the third law in
its strong form. The usual model provided to illustrate how
black hole mechanics does not satisfy the third law is given
by the Kerr-Newman black hole. It is in fact straightforward
to check that the parameters in the Kerr-Newman geometry
can be varied in a way such that $T$ goes to zero as $S$
remains a finite function of the charge and the angular momentum,
thus violating the `strong' form of the third law
\cite{Racz}.
This behaviour has been in turn interpreted as an
indication that the extremal solutions cannot be attained
by any finite physical process
\cite{israel},
or that a phase transition occurs when the extremal
values are reached and that under these regimes matter
behaves in an anti-thermodynamical way
\cite{Martellini},
or that a revision of the laws of black hole mechanics
from microscopic considerations in order to reproduce the
laws of thermodynamics is needed.

Given the fact that the extremal and super-extremal Kerr-Newman
spacetimes present a naked singularity, as the horizon
vanishes as the parameters reach the extremal value,
the analysis of the third law in this context has been
often associated to that of the CCC, and the validity
of some form of the third law has been suggested at times
as an argument in favor of the Cosmic Censorship.
We had like to point out here how the discussion
regarding the third law in extremal Kerr-Newman spacetimes
and the validity of CCC cannot be associated to the
discussion on the validity of CCC in collapse models.

In fact, the third law of black hole mechanics
stands on a different footing when applied to dynamical
spacetimes describing collapse or to extremal Kerr-Newman
solutions, just
as much as naked singularities arising in collapse models
stand on a different ground from those appearing in
stationary and axially symmetric spacetimes.
Generally in the static, non charged case, $T$ goes
as the inverse of $M$ and so, even for the Schwarzschild
black hole, it is impossible to reach $T=0$ with a finite
physical process, a statement that is in agreement
with the weak form of the third law.

In our case, neglecting rotation and charge, the
third law, in its `strong' form, still presents some issues
in the sense that the identification of a temperature
for a dynamical spacetime is a non-trivial task that
can lead to different conclusions
\cite{Nielsen}.

Approaching the singularity along the apparent horizon,
the surface gravity diverges as the area goes to a
constant value. This is not surprising, given the fact
that at the singularity the energy density is diverging
and since we have to remember that the singularity is
not part of the spacetime. Therefore, the values of $S$
and $T$ at the singularity must be taken only in a
limiting sense, or it would be correct to say that these
are actually not defined there. Nevertheless this behaviour
might indicate the necessity to find a more suitable definition
for $S$ or $T$ on the apparent horizon as it approaches
the singularity. For example, in the black hole case we
could set the temperature as the temperature of the event
horizon, as seen at spatial infinity, thus neglecting
the contribution to the entropy given by the apparent horizon,
that diverges as the singularity is approached but cannot
be seen by any outside observer.

On the other hand, in the collapse models described
above, the `weak' formulation of the third law remains
valid in a similar way as it is valid for the Schwarzschild
case, since for these models the surface gravity goes
to zero as $R$ goes to infinity, which corresponds
to the statement of physical unattainability.

\section{Concluding remarks}\label{discussion}

We discussed here a generic
gravitational collapse going either to a black hole or to a
naked singularity final state, and we examined the validity
of the laws of black hole thermodynamics in such a
dynamical scenario.

We note that in the Einstein theory, naked
singularities arise in a variety of ways:

\begin{enumerate}
  \item In extremal vacuum spacetimes such as
Kerr-Newman or Reissner-Nordstrom.
  \item In static and stationary axially symmetric
vacuum spacetimes such as the Zipoy-Voorhees and the
Tomimatsu-Sato metrics.
  \item In dynamically evolving spacetimes such
as the Lemaitre-Tolman-Bondi models.
\end{enumerate}

The first category above is that of the naked singularities
that appear in the extremal Kerr-Newman and Reissner-Nordstrom
spacetimes. These are not dynamical spacetimes, and the
naked singularity can be obtained by setting `ad-hoc'
certain values of the key parameters describing the solution.
Therefore the procedure by which such naked singularities
are achieved, starting from a non-extremal black hole,
is not necessarily a physical dynamical process. It is essentially
the variation of the basic parameters of the model (namely,
the charge and the angular momentum) in a stationary spacetime.
Hence, although it can be argued that the occurrence of
such singularities could be the endstate of some physical
processes (by which the above mentioned parameters can be varied),
the procedure by which they are obtained is not in general
the result of the dynamical evolution of the system
as per the Einstein equations.

Therefore, the attainability of such naked singularities
starting from a covered non-extremal black hole is a matter
of much discussion and debate at present
\cite{Liberati}.
There are some claims that suggest that these
naked singularities would violate the laws of black hole
thermodynamics, and therefore, accepting the analogy between
gravity and thermodynamics as a real feature of physics
would imply that these singularities cannot occur in the
physical universe. In this sense, this point of view links
the validity of the Cosmic Censorship Conjecture to the
laws of black hole thermodynamics.

The second category includes naked singularities
that appear in static or stationary axially symmetric
spacetimes. Again these are not dynamical spacetimes and the
occurrence of singularities is a direct consequence of the
chosen family of vacuum solutions.
Departing from spherical symmetry immediately
destroys the horizon structure, giving rise to a naked
singularity. In some cases it is possible to link such
spacetimes to spherical ones, for example, as in the
Zipoy-Voorhees solution, also known as the gamma metric,
or in the Tomimatsu-Sato metric, therefore obtaining a
procedure to move from spherical symmetry to axial symmetry
by the variation of one deformation parameter. Again, it
has been suggested that once the parameter is changed as
to depart from spherical symmetry, the thermodynamical
properties of the horizon are lost. This is not surprising
since the coordinates at which the horizon was located
become singular and do not any longer describe a portion
of the manifold. Nevertheless, once again, the variation of
the deformation parameter is not a dynamical process as
it can occur in the real universe. A similar situation
in reality would require a dynamically evolving spacetime
and the solution of Einstein equations for that case,
a scenario that is still faraway from our understanding
today.

The third category, which is the one we have investigated
here, describes singularities arising in dynamical spacetimes
evolving from some regular set of initial data. This is the
case of the well-known naked singularities in gravitational
collapse. For example, in the Lemaitre-Tolman-Bondi models the
presence of inhomogeneities in the collapsing dust cloud
can be enough to cause the collapse to end in a naked singularity.
It must be stated though
that naked singularities arising in many of the collapse models
are naked only for a `brief' period of time. The presence
of pressures can alter the usual black hole formation picture,
thus uncovering a portion of the singularity curve, but it
cannot undress the singularity at all times \cite{Mandar}.

We have shown here that the thermodynamical analysis of
the apparent horizon in such cases as the third category above,
is entirely possible and
that it does not violate the laws of black hole thermodynamics
as formulated for dynamical horizons. Therefore, a possible
proof of the validity of the connection between thermodynamics
and gravity on a general level would not necessarily rule out the
occurrence of such singularities, whether hidden within
a black hole or visible to external observers in the
universe.

We emphasize again that singularities cannot be
treated as part of the spacetime. Therefore one should
study the thermodynamical behaviour and laws only on
appropriate surfaces such as the apparent horizon and
not on the singularity itself, though one can approach
the singularity as close as one wishes.
We also note that, considering that naked singularities
do occur quite generally in Einstein's gravity, both in static
or stationary situations as in the dynamical cases,
it may also be the case that if they
are not consistent in some situations with the thermodynamical
behaviour, then gravity may not be equivalent to
thermodynamics in a global and fully general sense.

\end{document}